\documentclass[12pt]{article}
\newlength{\pubnumber} 

\catcode`\@=11
\@addtoreset{equation}{section}

\def\section{\@startsection{section}{1}{\z@}{3.5ex plus 1ex minus .2ex}
 {2.3ex plus .2ex}{\large\bf}}
\def\subsection{\@startsection{subsection}{2}{\z@}{2.3ex plus .2ex}
 {2.3ex plus .2ex}{\bf}}

\begin{document}

\begin{titlepage}
\samepage{
\setcounter{page}{0}
\rightline{CPHT-RR048.0708}
\rightline{April 2008}
\vfill
\begin{center}
    {\Large \bf Is SUSY Natural?\footnote{Invited article to appear 
in the New Journal of Physics, 
Special Focus Issue
``Particle Physics at the TeV Scale''.}} 
\vfill
\vspace{.13in}
   {\large
      Keith R. Dienes,$^1$\footnote{E-mail address:  dienes@physics.arizona.edu}~ 
    Michael Lennek,$^{1,2}$\footnote{E-mail address:  mlennek@cpht.polytechnique.fr}~
     David S\'en\'echal,$^3$\footnote{E-mail address:  david.senechal@usherbrooke.ca}\\
  Vaibhav Wasnik$^1$\footnote{E-mail address:  wasnik@physics.arizona.edu} \\}
\vspace{.18in}
 {\it  $^1\,$Department of Physics, University of Arizona, Tucson, AZ  85721  USA\\}
 {\it  $^2\,$Centre de Physique Th\'eorique, Ecole Polytechnique, \\ 
            CNRS, F-91128 Palaiseau, France\\}
 {\it  $^3\,$D\'epartement de Physique, Universit\'e de Sherbrooke, \\ 
            Sherbrooke, Qu\'ebec J1K 2R1  Canada\\}
\end{center}
\vfill
\begin{abstract}
  {\rm  
       Spacetime supersymmetry is widely believed to play an important role in most
       fundamental theories of physics, and is usually invoked in order to address
       problems of naturalness.
       In this paper, we examine the question
       of whether supersymmetry itself is ``natural'' 
       ({\it i.e.}\/, likely to exist as a fundamental component of nature at
        high energy scales).  Our approach to answering this question is 
       based on a statistical examination of the heterotic string landscape, and 
       our conclusion is that supersymmetry is an exceedingly rare phenomenon. 
       We also find that the likelihood of supersymmetry appearing at the 
         string scale is dependent on the gauge symmetries present at 
         the string scale,
         with certain gauge groups strongly favoring the appearance of
            ${\cal N}{=}1$ supersymmetry and others not.  
      This article summarizes several recent papers, yet also
         contains some new results. 
      In particular, one new result is that the heterotic landscape appears to favor
     either the non-supersymmetric Standard Model or an ${\cal N}{=}1$~SUSY GUT
      gauge group at the string scale;  by contrast, the opposite outcomes (namely
        the MSSM or a non-supersymmetric GUT)
       are significantly disfavored.
  }
\end{abstract}
\vfill}
\end{titlepage}

\setcounter{footnote}{0}

\def\beq{\begin{equation}}
\def\eeq{\end{equation}}
\def\beqn{\begin{eqnarray}}
\def\eeqn{\end{eqnarray}}

\def\calO{{\cal O}}
\def\calE{{\cal E}}
\def\calT{{\cal T}}
\def\calM{{\cal M}}
\def\calF{{\cal F}}
\def\calY{{\cal Y}}
\def\calV{{\cal V}}
\def\calN{{\cal N}}
\def\ibar{{\overline{\i}}}
\def\qbar{{\overline{q}}}
\def\mm{{\tilde m}}
\def\ahat{{\hat a}}
\def\nn{{\tilde n}}
\def\rep#1{{\bf {#1}}}
\def\ie{{\it i.e.}\/}
\def\eg{{\it e.g.}\/}

\def\Str{{{\rm Str}\,}}
\def\bone{{\bf 1}}

\def\thetai{{\vartheta_i}}
\def\thetaj{{\vartheta_j}}
\def\thetak{{\vartheta_k}}
\def\thetaibar{\overline{\vartheta_i}}
\def\thetajbar{\overline{\vartheta_j}}
\def\thetakbar{\overline{\vartheta_k}}
\def\etainv{{\overline{\eta}}}

\def\modinvmeasure{{  {{{\rm d}^2\tau}\over{\tautwo^2} }}}
\def\qbar{{  \overline{q} }}
\def\ahat{{ \hat a }}

\newcommand{\newc}{\newcommand}
\newc{\gsim}{\lower.7ex\hbox{$\;\stackrel{\textstyle>}{\sim}\;$}}
\newc{\lsim}{\lower.7ex\hbox{$\;\stackrel{\textstyle<}{\sim}\;$}}
\hyphenation{su-per-sym-met-ric non-su-per-sym-met-ric}
\hyphenation{space-time-super-sym-met-ric}
\hyphenation{mod-u-lar mod-u-lar--in-var-i-ant}


\def\inbar{\,\vrule height1.5ex width.4pt depth0pt}

\def\IC{\relax\hbox{$\inbar\kern-.3em{\rm C}$}}
\def\IQ{\relax\hbox{$\inbar\kern-.3em{\rm Q}$}}
\def\IR{\relax{\rm I\kern-.18em R}}
 \font\cmss=cmss10 \font\cmsss=cmss10 at 7pt
\def\IZ{\relax\ifmmode\mathchoice
 {\hbox{\cmss Z\kern-.4em Z}}{\hbox{\cmss Z\kern-.4em Z}}
 {\lower.9pt\hbox{\cmsss Z\kern-.4em Z}}
 {\lower1.2pt\hbox{\cmsss Z\kern-.4em Z}}\else{\cmss Z\kern-.4em Z}\fi}

\long\def\@caption#1[#2]#3{\par\addcontentsline{\csname
  ext@#1\endcsname}{#1}{\protect\numberline{\csname
  the#1\endcsname}{\ignorespaces #2}}\begingroup
    \small
    \@parboxrestore
    \@makecaption{\csname fnum@#1\endcsname}{\ignorespaces #3}\par
  \endgroup}
\catcode`@=12

\section{Introduction}
\setcounter{footnote}{0}
\label{intro}

Most theoretical frameworks for physics beyond the Standard Model involve
the introduction of supersymmetry (SUSY), and there are many reasons why this is so.
First, supersymmetry solves the technical gauge hierarchy problem.
Second, supersymmetry provides a dynamical mechanism for triggering electroweak 
symmetry breaking.  Third, supersymmetry improves the accuracy of gauge coupling
unification, and fourth, it provides a dark matter candidate.
As a result, supersymmetry is truly ubiquitous in particle physics,
appearing virtually everywhere  --- except the data.

However, through the years, lots of competing or alternative theories have been proposed.
Some involve large extra dimensions and some involve small extra dimensions.
Others do not explicitly involve extra dimensions at all, yet contain new, strongly coupled 
sectors.  Likewise, especially over the past decade, the substance of phenomenological 
model-building has changed dramatically.  Indeed, it is now quite common that a talk
introducing
a new particle-physics
scenario will begin with a litany of sequential assumptions 
that would have sounded increasingly fantastic to the ear of a physicist a mere
decade ago.
We are made of open strings.
And we live on a brane.
And the brane lives in extra dimensions.
And the brane is wrapped and intersects other branes.
And the extra dimensions are warped.
And the warping is severe and forms a throat.
And the brane is falling into a throat.
And so forth, and so on.
Indeed, such scenarios now often form the backbone of cutting-edge model-building.

Admittedly, all of this may sound highly unnatural, and it is excusable to yearn 
for the simpler days of the MSSM and their cousins, the SUSY GUTs.
But is SUSY itself truly natural?
What does it mean to be ``natural'', anyway?

There are many different notions of naturalness that have appeared in the literature.
For example, Dirac naturalness
stipulates that an effective field theory (EFT) is natural if the dimensionless coefficients for all operators
are $\sim{\cal O}(1)$ --- \ie, no exceedingly small or large numbers are allowed.
In this sense, the large electroweak gauge hierarchy is unnatural, which is one of the biggest motivations
for supersymmetry.
Another notion is 't~Hooft naturalness:  even if such a number is small, it can be viewed as ``natural''
if its smallness is protected by a nearly unbroken symmetry. 

But neither of these addresses the question of whether a theory, even if ``natural''
in the above sense, is {\it likely}\/ to be right.
How {\it likely}\/ is SUSY to be the correct theory?

The word {\it likely}\/ often causes us to shudder.
Indeed, even though we constantly judge theories this way, we don't say this
word aloud because the question of theoretical likelihoods seems 
more philosophical than scientific, especially when we have no data upon
which to base our assessments.
How likely relative to {\it what}\/?
To all other theories that one can imagine?
And who is doing the imagining?
One might get very different responses depending on the identity of the
unlucky proponent.
Ultimately, we seem to be faced with a dead-end question.
 {\it How can one compare the likelihood of one theory against another?}

String theory provides a framework in which 
this question can be addressed in a mathematical way.
This is because string theory provides a large set of possible
vacuum solutions (``vacua'', or string ``models'', collectively called
``the landscape''), each of which corresponds to a different
alternative universe with different physical laws.
In this context, we can then place our likelihood question on firmer
statistical footing, as originally advocated in Ref.~\cite{Douglas}.  
 {\it In the landscape of possible string solutions, how many of these solutions
are supersymmetric?  Is SUSY ``natural'' on this landscape, or relatively rare?}

This is the subject of this article.

\section{Our Study:  the Heterotic Landscape}
\setcounter{footnote}{0}

In order to approach this question in a somewhat tractable way, we shall focus on the landscape
associated with perturbative four-dimensional heterotic strings which are realizable
using free-field constructions (such as those based on free worldsheet bosons or fermions).
Our investigation will then focus on the fundamental question of determining the probability
that such theories exhibit spacetime supersymmetry at the string scale.
Note that this is different from previous
discussions~\cite{Dine} in which a certain intrinsically supersymmetric
framework~\cite{KKLT} for the high-scale 
theory is assumed, and in which one asks about the likelihood that dynamical
supersymmetry breaking is subsequently generated near the electroweak scale.

In a recent paper~\cite{Paper1}, we conducted a random exploration of the landscape associated with such models,
using the free-fermionic construction method~\cite{ffmethod}.
In this construction, each model can be described in terms of its left- and right-moving worldsheet conformal field 
theories.  These conformal field theories consist of tensor products of non-interacting, free, complex fermionic fields.  
Different models are then achieved in this method by varying the boundary conditions of these worldsheet fermionic fields around 
the two non-contractible loops of the worldsheet torus.  The sets of allowed boundary conditions for the fermionic 
fields are restricted by numerous string self-consistency conditions which also must be applied.  These 
self-consistency conditions guarantee that the string partition function can be viewed as
the trace over a Fock space corresponding to a self-consistent string model.
Other perturbative self-consistency constraints (such as conformal and modular invariance) are also imposed.

This study was similar to earlier studies performed in Refs.~\cite{Senechal,dienes},
and resulted in a data set of approximately $10^7$ distinct self-consistent four-dimensional
heterotic string models, each of which is tachyon-free and hence stable at tree level.
To the best of our knowledge, this is the largest set of distinct heterotic string models ever
constructed.  
However, this study has some important limitations.
For example, this study examined string models for which all real worldsheet fermions could 
be grouped together to form complex worldsheet fermions.  This effectively reduces the number of 
consistent sets of fermion worldsheet boundary conditions, but allows for the utilization of many 
time-saving computational algorithms.  The major phenomenological consequence of this restriction is 
that these string models have gauge groups with a fixed rank (namely twenty-two in four dimensions).

In our study, two models were considered distinct if their spacetime phenomenologies differ in some way.
For example, two models were considered distinct if they differ in their amounts
of unbroken supersymmetry, or in their gauge group or massless particle spectrum.  Moreover,
using data gathered from our sample set of models, a novel technique~\cite{Prob} 
was utilized in order to extract stable statistical results which do not vary as a function of sample
size.  As discussed in Ref.~\cite{Prob}, this method transcends a mere statistical analysis of the models 
in our limited sample set, and yields results which are likely to be indicative of the corresponding 
landscape associated with such models as a whole.
The techniques by which such results are extracted is discussed in Ref.~\cite{Prob}.

\section{Results}
\setcounter{footnote}{0}
\label{PrevRes}

In this section, we shall describe the results of our study.
As indicated above, the statistical methodology utilized in extracting these results is discussed
in Ref.~\cite{Prob}, and a full discussion of the results of Sects.~3.1 and 3.2 can be found in Ref.~\cite{Paper1}.

\subsection{Prevalence of SUSY on the heterotic landscape}

One of the first questions addressed in our study concerned the prevalence of spacetime supersymmetry 
in the heterotic landscape.  
In other words, what percentage of the heterotic string landscape
consists of string models exhibiting an unbroken spacetime supersymmetry at the string scale? 
Clearly, answering this question is the first step towards addressing the overall issue of the naturalness of supersymmetry.  

Our results~\cite{Paper1} are reproduced in Table~\ref{SUSYLand}.
Note that in quoting these results, we are explicitly focusing on only that portion of the heterotic string
landscape which is stable at tree level. 
In other words, we are explicitly disregarding those non-supersymmetric
portions of the landscape (which amount to approximately 32.1\% of the total)
for which tachyonic states exist at tree level.

\begin{table}[htb]
\begin{center}
\begin{tabular}{||c|c||}
\hline
\hline
SUSY class & \% of heterotic landscape\\
\hline
${\cal N}{=}0$ (tachyon-free) & $68.48$\\
${\cal N}{=}1$ & $30.78$\\
${\cal N}{=}2$  & $0.74$ \\
${\cal N}{=}4$  & $0.0044$ \\
\hline
\hline
\end{tabular}
\end{center}
\caption{
   Classification of the four-dimensional heterotic landscape 
 as function of the number of unbroken spacetime supersymmetries.  We are explicitly 
    focusing on string models which are stable (and thus tachyon-free) at tree level.}
\label{SUSYLand}
\end{table}

Table~\ref{SUSYLand} represents our final partitioning of the tree-level four-dimensional
heterotic landscape according to its degree of supersymmetry.
There are several rather striking facts which are evident from these results.
 \begin{itemize}
\item{}  First, we see that more than half of the stable heterotic landscape is
             non-supersymmetric and yet tachyon-free.  Indeed, this proportion
            remains near half of the total even when the non-supersymmetric tachyonic
            models are included.
\item{}  Second,  we see that the supersymmetric portion of the heterotic
           landscape appears to account for less than one-third of
           the full tachyon-free four-dimensional heterotic landscape.
\item{}  Finally, models exhibiting extended ($\calN\geq 2$) supersymmetries are exceedingly
           rare, together representing less than one percent of the full landscape.
\end{itemize}

Of course, we stress once again that these results hold only for the {\it tree-level}\/ landscape,
\ie, models which are stable at tree level only.
 {\it A priori}\/, it is not clear whether these results would persist after full moduli stabilization.
However, it seems likely that they would, given that most modern methods of moduli stabilization
(fluxes, superpotentials, {\it etc}\/.) tend to further break (rather than restore) spacetime supersymmetry.
Indeed, under these assumptions,
these results then lead to a number of interesting conclusions.

The first conclusion is that the properties of the tachyon-free heterotic
landscape as a whole are statistically
dominated by the properties of string models which do {\it not}\/ have spacetime
supersymmetry.  Indeed, the $\calN{=} 0$ string models account for nearly three-quarters
of this portion of the heterotic
string landscape.  The fact that the $\calN{=} 0$ string models dominate the tachyon-free portion of the
landscape suggests that breaking supersymmetry without introducing tachyons
is actually {\it favored}\/ over
preserving supersymmetry for this portion of the landscape.
Indeed, we expect this result to hold even after full moduli stabilization
(as has been argued within the context of Type~I strings~\cite{Silverstein}),
unless an unbroken supersymmetry is somehow restored by stabilization.

The second conclusion which can be drawn from these results is that the
supersymmetric portion of the landscape is almost completely comprised of
$\calN{=}  1$ string models.  Indeed, only  $2.4 \%$ of the supersymmetric
portion of the heterotic landscape has more than $\calN{=} 1$ supersymmetry.  This suggests that
the correlations present for the supersymmetric portion of the landscape
can be interpreted as
the statistical correlations within the $\calN{=} 1 $ string models,
 with the $\calN{=} 2$ correlations representing a correction at
the level of $2\%$ and the $\calN{=} 4$ correlations representing a nearly negligible correction.

In fact, the SUSY faction of the full string landscape may be even smaller than
quoted here.  One reason is that free-field string constructions (such as we are
employing here) probably tend to artificially favor models with unbroken supersymmetry.
Second, even when stabilized string models nevertheless exhibit spacetime SUSY 
at the string scale, there remains the difficult quesion of determining the statistical
likelihood that this SUSY will survive all the
way down to the electroweak scale~\cite{Dine}.  At the very least, this should restrict
the number of string models leading to weak-scale SUSY still further. 

Thus, we conclude that weak-scale SUSY is rather {\it unnatural}\/ from a string landscape
perspective.  On the one hand, this result shifts the burden of proof onto the SUSY enthusiasts,
which represents a dramatic reframing of the underlying question of whether SUSY should exist
at or above electroweak scale.  But the conclusion that weak-scale SUSY is unnatural should not
necessarily be viewed as a problem for string phenomenology.  In fact, this result might even
be considered good news:  it implies that we will actually learn something about string theory
and its preferred compactifications if/when weak-scale supersymmetry is actually discovered
in upcoming collider experiments!

\subsection{Correlations between SUSY and gauge groups}
\setcounter{footnote}{0}

We shall now examine the effects of supersymmetry on the probability of realizing different gauge group factors 
in this landscape.  What percentage of string models with a given level of supersymmetry will contain a given gauge group
factor amongst its unbroken gauge group at the string scale?
We shall also be interested in knowing the likelihood of realizing different gauge group factors for the full landscape as a whole.  

Our results~\cite{Paper1} are presented in Table~\ref{ApGroup}.  
As can clearly be seen, supersymmetry has a profound effect upon the prevalence of different gauge group factors.
Moreover, even independently of SUSY, there are some general trends which emerge from these results.
These trends include:
\begin{itemize}
\item{}  A preference for $SU(n+1)$ over $SO(2n)$ groups for each rank $n$.
         Even though these two groups have the same rank, it seems that
         $SU$ groups are more common than the $SO$ groups for all levels of
         supersymmetry.
\item{}   Groups with smaller rank are much more common
            than groups with larger rank.  Once again, this also appears to hold
         for all levels of supersymmetry.
\item{}   Finally, the gauge-group factors comprising Standard-Model gauge group
           $G_{\rm SM}\equiv SU(3)\times SU(2)\times U(1)$
          are particularly common, much more so than those of any of its grand-unified extensions.
\end{itemize}

\begin{table}[ht]
\begin{center}
\begin{tabular}{||c||r|r|r|r||c||}
\hline
\hline
 {\small gauge group} & ${\cal N}{=}0$&~${\cal N}{=}1$~&~${\cal N}{=}2$~&~${\cal N}{=}4$~&{\small ~full landscape~}\\
\hline
\hline
$U_1$ &99.9 & 94.5 & 68.4 &89.6& 98.0\\
\hline
$SU_2$ & 62.46 & 97.4 & 64.3 & 60.9 & 73.2\\
\hline
$SU_3$ & 99.3 & 98.0  & 93.0  & 45.1 & 98.9\\
\hline
$SU_4$& 14.46 & 30.0 & 39.0  & 53.5 & 19.4\\
\hline
$SU_5$&16.78 & 43.5 & 66.3 & 33.8 & 25.4\\
\hline
$SU_{>5}$ &0.185 &1.7 & 10.6 &73.0 &0.73\\
\hline
$SO_8$& 0.482 & 1.6  & 6.2  & 21.1 & 0.87\\
\hline
$SO_{10}$&0.084  & 0.2  & 1.6  & 18.7 & 0.13\\
\hline
$SO_{>10}$ & ~$0.005$ &0.038  & 0.77 & 7.5 & 0.021\\
\hline
$E_{6,7,8} $& ~$0.0003$ & 0.03 & 0.16 & 11.5 & 0.011\\
\hline
\hline
\end{tabular}
\end{center}
\caption{The percentage of heterotic string models exhibiting specific gauge group factors as functions of 
their spacetime supersymmetry.  Here $SU_{>5}$ and $SO_{>10}$ collectively indicate gauge groups $SU(n)$ 
and $SO(2n)$ for any $n>5$, while ${\cal N}$ refers to the number of unbroken supersymmetries at the string scale.  
Note that the $\calN{=}0$ models are all tachyon-free.  The rightmost column of this table is derived from the
other columns using the landscape weightings in Table~\protect\ref{SUSYLand}.}
\label{ApGroup}
\end{table}

\subsection{SUSY naturalness}
\setcounter{footnote}{0}
\label{Particle}

At this point, we have presented both the unrestricted probability of finding different levels of supersymmetry
on the heterotic landscape,  and the restricted probability of finding different gauge group factors when 
a certain degree of supersymmetry is assumed.
However, a more useful quantity might be the ``inverse'' of this last probability, 
namely the probability of finding different levels of supersymmetry given a specific gauge group factor.  
This would give an indication of how ``natural'' each level of supersymmetry is for each
different possible gauge group factor.  

In order to derive these probabilities, we can utilize the results presented above.
First, let us recall some basic elements of probability theory.  
The results in the first four columns of Table~\ref{ApGroup} are necessarily conditional probabilities:  each number 
in these columns represents the probability of finding a specific gauge group factor {\it given}\/ a certain level of supersymmetry.  
Thus, if $A$ represents the occurrence of a specific gauge group factor and $B$ represents the occurrence of a specific 
level of supersymmetry, then the results presented in Table~\ref{ApGroup} are
all of the form
\beq
              P(A\mid B) ~\equiv~ \frac{P(A \cap B)}{P(B)}~.
\eeq
Of course, what we now seek is not $P(A\mid B)$, but the ``inverse'' $P(B\mid A)$.
In general, the relationship between $P(B\mid A)$ and $P(A \mid B)$ is 
given as
\beq
                 P(B \mid A) ~=~ \frac{P(B)}{P(A)} ~P(A\mid B)~.
\label{Bayes}
\eeq
Fortunately, all of the probabilities needed within the right side of Eq.~(\ref{Bayes}) are present in Tables~\ref{SUSYLand} and \ref{ApGroup}.
Specifically, $P(A)$ is given within the rightmost column of Table~\ref{ApGroup}, while
$P(B)$ is given in Table~\ref{SUSYLand} and $P(A\mid B)$ is given in the rest of Table~\ref{ApGroup}.
Thus, given the presence of specific gauge group factor at the string scale,
we can now determine the corresponding probability of finding different levels of unbroken supersymmetry at the 
string scale.  Our results are presented in Table~\ref{Gold}.
Note that in calculating the results in Table~\ref{Gold}, we have retained more significant
digits than are explicitly shown in Tables~\ref{SUSYLand} or \ref{ApGroup}.

\begin{table}
\begin{center}
\begin{tabular}{||c||c|c|c|c|c|c|c|c|c|c||}
\hline
\hline
 {\small SUSY} & $U_1$ & $SU_2$& $SU_3$& $SU_4$& $SU_5$& $SU_{>5}$& $SO_8$& $SO_{10}$& $SO_{>10}$& $E_{6,7,8}$\\
\hline
$\calN = 0$& $69.80$ & $58.41$& 68.79 & 50.98 & 45.29 & 17.33 & 37.98 & 43.68 & 16.21 & 1.85\\
\hline
$\calN = 1$& $29.68$ & 40.94 & 30.51 & 47.53 & 52.78 & 71.56 & 56.66 & 46.75 & 55.38 & 83.00\\
\hline
$\calN = 2$& $0.51$ & 0.65 & 0.69 & 1.48 & 1.92 & 10.65 & 5.25 & 8.95 & 26.84 & 10.59\\
\hline
$\calN = 4$& $0.004$ & $0.002$ & $0.002$& $0.012$ & $0.006$& $0.44$& $0.11$& $0.63$& $1.57$& $4.57$\\
\hline
\hline
\end{tabular}
\end{center}
\caption{The percentage of string models with different levels of supersymmetry as a function of different gauge group factors.  
Thus, if we know that a given string model 
gives rise to a specific gauge group factor at the string scale,
this table lists the  corresponding probabilities that this model will have various levels of unbroken supersymmetry.
This table can therefore be viewed as the ``inverse'' of Table~\ref{ApGroup}.}
\label{Gold}
\end{table}

By comparing these probabilities to the probabilities given in Table~\ref{SUSYLand}, it is possible 
to determine which gauge group factors tend to favor different levels of supersymmetry {\it beyond their 
expected representations on the landscape as a whole}\/.
For example, we see from Table~\ref{SUSYLand} that $68.48\%$ of the stable heterotic landscape is non-supersymmetric.
Thus, if a given gauge group factor is associated with models of which fewer than $68.48\%$ are non-supersymmetric,
then this gauge group factor can be said to preferentially {\it favor}\/ supersymmetry. 

These results have a number of dramatic implications.

\begin{itemize}
\item  First, we observe that gauge group factors with large rank (greater than four) 
             actually {\it favor}\/ the appearance of unbroken supersymmetry.
\item  Second,  we observe that the gauge group factors which comprise the Standard Model gauge group 
            do {\it not}\/ generally favor supersymmetry.
\item  Finally, we see that the $SU(n)$ gauge group factors (with $n>5$) 
          and the exceptional gauge groups $E_{6,7,8}$ overwhelmingly favor $\calN = 1$ supersymmetry.
          This preference is substantial, resulting from the combined effects of the individual
          probabilities contributing to Eq.~(\ref{Bayes}).
\end{itemize}
We thus conclude that the heterotic string landscape appears to favor either the non-supersymmetric Standard Model
gauge group or an ${\cal N}=1$ SUSY GUT gauge group at the string scale.  However, the opposite outcomes 
(namely the MSSM or a non-SUSY GUT gauge group) are significantly disfavored.

One important caveat is that the gauge group factors presented in Table~\ref{Gold} do {\it not}\/ generally specify 
the gauge group fully.  Indeed, these gauge group factors could be part of either a hidden sector or the visible sector.  
However, the gauge group factors listed in these tables are necessarily among those which are
explicitly present in the full gauge group of the string model at the string scale.

\section{Discussion}
\label{Conc}

In this paper, we have presented results concerning the prevalence of spacetime supersymmetry at the string scale
and its possible statistical correlations with the unbroken gauge group which might also appear at the string scale.
Since these two quantities (spacetime supersymmetry and unbroken gauge group) are completely independent
in an ordinary quantum field theory based on point particles, these sorts of correlations represent true predictions
of string theory and thereby provide one possible route towards answering the question as to whether supersymmetry
is truly ``natural'' as a component of Beyond-the-Standard-Model physics.

As we have seen, the results of this calculation show that spacetime supersymmetry is {\it not}\/ 
generically a feature of the low-energy limit of string theory.  However, spacetime supersymmetry 
is actually statistically favored for certain gauge group factors.  

There are some inherent limitations to these results which must continually be borne in mind.  
For example, these string models are generally unstable:  
the non-supersymmetric models generically have non-zero dilaton tadpoles, and the supersymmetric 
string models have flat directions.  
Thus, one might think that these results might change if only fully-stabilized models are considered.  
Unfortunately, no fully stable perturbative heterotic string models have ever been constructed.  
One could even argue that if the non-supersymmetry string models were required to 
be as ``stable'' as the supersymmetric string models (\eg, only have some finite number of flat directions), 
then these results would also change.  However, at this point in time, the string models considered in 
this study are state-of-the-art and are as stable as many of the other classes of string models which have been 
considered in other statistical studies.  

Another issue facing this study concerns the extent to which these sorts of statistical correlations can be
trusted, given that the full heterotic landscape has not been surveyed.  
However, this issue has been discussed in Ref.~\cite{Prob}, and the methods developed there have been
used in order to extract each of the results quoted here.  Thus, to the best of our knowledge,
the results quoted here are independent of the size of our sample of heterotic string models,
and thus should persist across the landscape as a whole.

Finally, one could argue
that the construction method utilized in this study necessarily only probes certain sections of 
the heterotic landscape.  While this is true, these sections are the ones most likely to contain string
models realizing non-abelian gauge symmetries.
As such, these are the sections most likely to give rise to
non-trivial low energy phenomenologies.

The interpretation of these results is also open to some debate.  
The probability-based definition of naturalness used in this paper is not the traditional one,
and may only hold relevance for a landscape study such as the one we are performing.  
However, this definition of naturalness has the advantage of being applicable in a wide variety of
contexts, and does not resort to any aesthetic or theoretical prejudices concerning the parameters that
appear in effective Lagrangians.
As such, probability-based definitions of naturalness may have
inherent advantages over other definitions.

There are several extensions to these results which are currently under investigation.
For example, we would like to understand how the presence of supersymmetry affects the
statistical appearance of the entire composite Standard-Model gauge group
$G_{\rm SM}\equiv SU(3)\times SU(2)\times U(1)$, and not merely
the appearance of its individual factors.  We would also like to understand how the presence or
absence of supersymmetry affects other features which are equally important
for the overall architecture of the Standard Model:  these include the appearance
of three chiral generations of quarks and leptons, along with a potentially correct set of gauge
couplings and Yukawa couplings.  This work will be reported elsewhere~\cite{toappear}.

\section*{Acknowledgments}

The work of KRD, ML, and VW
was supported in part by the 
U.S. Department of Energy under Grant~DE-FG02-04ER-41298, 
by the 
U.S. National Science Foundation under Grant PHY/0301998,
and by a Research Innovation Award from
Research Corporation.
Since September 2007, the work of ML has been partially supported 
by ANR grant ANR-05-BLAN-0079-02,
RTN contracts MRTN-CT-2004-005104 and
MRTN-CT-2004-503369, CNRS PICS \#~2530, 3059 and 3747,
and by a European Union Excellence Grant MEXT-CT-2003-509661.
We are happy to thank M.~Trapletti for discussions,
and P.~Fox and S.~Raby for suggestions leading to the
new results reported in Sect.~3.3.

\bigskip
\bibliographystyle{unsrt}

\end{document}